\documentclass[12pt]{article} 
\usepackage[dvips]{graphics}
\title{ Rotation Eigenvectors and Spin 1/2 }  
\author{{\it Richard Shurtleff~}\thanks{affiliation and mailing 
address: Department of Mathematics and Applied Sciences, 
Wentworth Institute of Technology, 550 Huntington Avenue, 
Boston, MA, USA, ZIP 02115, telephone number: (617) 989-4338, fax 
number: (617) 989-4591 , page: www.public.wit.edu/faculty/shurtleffr/, e-mail address: shurtleff@wit.edu}} 
\date{March 28, 1999}
\begin{document} 
          
\maketitle               
			\begin{abstract}  

	It is an easily deduced fact that any four-component spin 1/2 state for a massive particle is a linear combination of pairs of two-component simultaneous rotation eigenstates, where `simultaneous' means the eigenspinors of a given pair share the same eigenvalue. The new work here constructs the reverse: Given pairs of simultaneous rotation eigenvectors, the properties of these pairs contains relationships that are equivalent to spin 1/2 single particle equations. Thus the needed aspects of space-time symmetry can be produced as special cases of more general properties already present in the rotation group. The exercise exploits the flexibility of the rotation group in three dimensions to deduce relativistic quantities in four dimensions.

1999 PACS number(s): 03.65.Fd; 03.65.Pm

			\end{abstract}
\pagebreak
\section{Introduction} 

	A subgroup of a group is less special and more general than the whole group. For example, all figures in a plane symmetric under rotations of 60 degrees are also symmetric under rotations of 120 degrees. But the reverse is not true, 120 degree symmetry does not imply 60 degree symmetry. So it is with the rotation group and the Lorentz group. In this paper we deduce spin 1/2 single particle states by constraining the more general properties of vectors in the rotation group. One expects this to be possible since the rotation group as a subgroup of the Lorentz group has more general properties than the whole group. 

	First consider spin 1/2 with the full Lorentz group. One may choose to write any spin 1/2 wave function for a single massive particle in any one of many equivalent representations. In a chiral representation the four-component wave function  contains one left-handed and one right-handed two-component spinor. A massive particle can be stopped and when at rest the two spinors are equal. Rotations act the same on both handed spinors, preserving the equality.

\begin{figure}[h] 	
\vspace{0in}
\hspace{0.75in}\includegraphics[0,0][360,180]{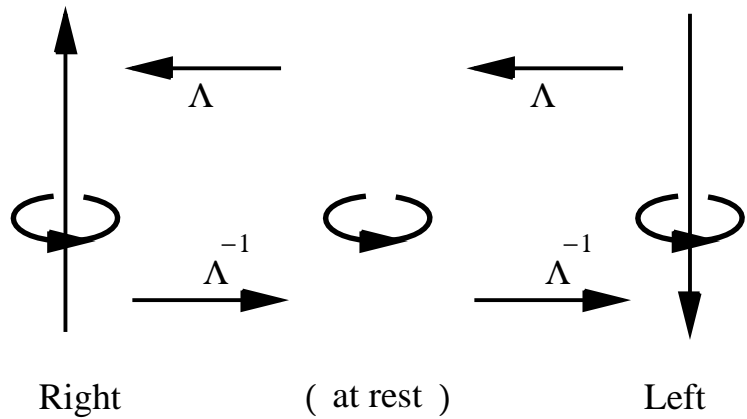}
\caption{A right spinor and a left spinor, each boosted from a spinor at rest. $\Lambda$ is the $2 \times 2$ matrix representing the boost from velocity $-\tanh u$ to 0 and from 0 to $+\tanh u$. Note the sketch is invariant under a spatial inversion, so the Left velocity (down arrow) is the Right velocity (up arrow) in a left-handed coordinate system. The spinor at rest is pre-selected to be an eigenvector of rotations about the boost direction, so all three spinors are eigenvectors with the same rotation eigenvalues.}
\end{figure}

	Being two-dimensional the two eigenspinors of a given rotation, one with plus eigenphase and one with negative eigenphase, serve as a basis for the eigenspinors of all rotations. It follows that, given a rest spin 1/2 wave function that is to be boosted in some direction, say z, we can first rewrite the left and right-handed spinors as combinations of z-axis rotation eigenspinors. The z boosts and z-axis rotations share eigenspinors, so the z boost applies numerical factors to the spinors. This makes a numerical factor relating the two eigenvectors, a factor dependent on speed with unity for speed zero. Since any two eigenspinors with the same eigenvalue are multiples, we conclude that the collection of all such pairs spans the space of spin 1/2 wave functions. 

	One way to interpret the Dirac equation for a free spin 1/2 particle involves boosting a left spinor to a right one and boosting the right spinor back to the left spinor, see Fig.~1. For details see Ref.~1-3. Boost matrices appear to be essential features in this interpretation of the equation.  This ends our consideration of spin 1/2 with the full Lorentz group.[4]

\begin{figure}[h] 	
\vspace{0in}
\hspace{0.75in}\includegraphics[0,0][360,180]{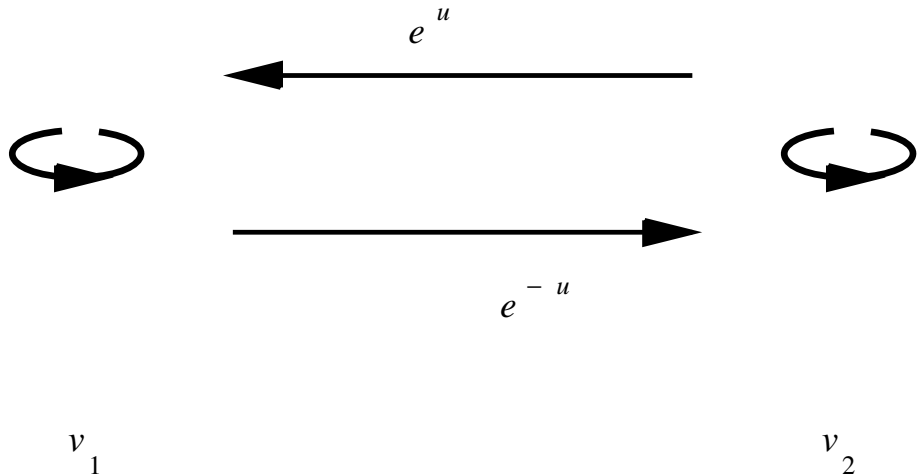}
\caption{Dismissing boosts and velocities, the vectors are still eigenvectors under rotations about the axis indicated in Fig.~1. With only two components, it follows that the eigenvectors must be related by numerical factors as illustrated here. In the text we rewrite the numerical factors $e^{u}$ and $e^{-u}$ as matrices to replace the $\Lambda^{2}$ and $\Lambda^{-2}$ of Fig.~1.}
\end{figure}

	In the rest of this paper we reverse the argument and show by construction how the properties of pairs of rotation eigenvectors suffice to produce spin 1/2 single particle equations. Thus we dispense with boosts, velocities and even space-time and attempt to base the Dirac equation on rotations in an abstract Euclidean space, see Fig.~2. 

	In Sec.~2, we consider the rotation group in an abstract three dimensional space. The work depends on two facts: Any two simultaneous eigenvectors are related by a numerical factor. And multiplication by either of two matrices, the unit matrix and one other, has no effect on the eigenvectors of the rotations about a given axis. Thus the numerical factor relating the two simultaneous eigenvectors is a special case of a more general matrix factor. Another special case produces the boost matrix in Fig.~1. Energy-momentum and left and right spinors can be then be identified and we show the pair of eigenvectors satisfies the Dirac equation in a momentum representation.

 	In Sec.~3 we search for space-time coordinates along with a suitable gradient form for the matrix derived in Sec.~2. First we note that the common eigenvalue of the two eigenvectors is a phase factor and the derivative of the phase factor with respect to phase gives the same phase factor back. Since the phase derivative is an operation on rotated eigenvectors that returns the original eigenvectors unchanged, we can introduce the phase derivative into the Dirac equation found in Sec.~2. Next, space-time coordinates are deduced by defining the space-time gradient in terms of the phase derivative, i.e. $\partial_{\mu}$ are functions of $d_{\theta}.$ The function is chosen to obtain the energy-momentum quantum operator in a coordinate representation. Lastly, it is shown that the Dirac equation in its first order partial differential version has been obtained. Linear differential equations obey a superposition principle; linear combinations of solutions are solutions. Thus any spin 1/2 wave function can be obtained in this way and the construction is complete. 

\section{Simultaneous eigenvectors} 

	The rotation group lacks the translations and boosts of the full Lorentz group. With the full Lorentz group, a boost can be represented by a matrix much like a rotation $R$. In this section we find a general matrix factor relating two eigenvectors of a given rotation $R.$ We then force the general matrix into the form of a boost. The result gives an equation suitable to relate left and right spinors. 

	Let the rotations act in an abstract three dimensional Euclidean space, which is not as yet related to any space in any reference frame of space-time. The rotation group in more-than-three dimensional Euclidean spaces will be considered elsewhere.

	Consider a two-dimensional representation of the rotation group in three dimensional Euclidean space. Let $v_{1}$ and $v_{2}$ be two eigenvectors of a rotation matrix $R$, 
\begin{equation} 
R v_{a} = e^{(i n^{k} \sigma^{k} \theta/2)} v_{a} = e^{i \theta/2} v_{a},
\end{equation}
where $a \in$ $\{1,2\}$, summation is assumed over $k \in$ $\{1,2,3 \},$ and $$\sigma^{1} = \pmatrix{0 && 1 \cr 1 && 0} \hspace{0.3in} \sigma^{2} = \pmatrix{ 0 && -i \cr i && 0} \hspace{0.3in} \sigma^{3} = \pmatrix{ 1 && 0 \cr 0 && -1} \hspace{0.3in} \sigma^{4} = \pmatrix{ 1 && 0 \cr 0 && 1}.$$
We added the unit $2 \times 2$ matrix to the list, calling it $\sigma^{4}$.

	Two eigenvectors $v^{+}$ and $v^{-}$, where $R v^{\pm}$ = $e^{\pm i \theta /2} v^{\pm}$, cannot be multiples of each other and therefore these two span the space of (two-component) vectors. It follows that $v_{1}$ and $v_{2}$ are multiples of $v^{+}$ and therefore of each other. Let the factor be $e^{u}$, $v_{1}$ = $e^{+u} v_{2}$ and $v_{2}$ = $e^{-u} v_{1}$, or more compactly,
\begin{equation} 
v_{a} = e^{\epsilon_{ab} u} v_{b} ,
\end{equation}
where $\epsilon_{12}$ = +1, and $\epsilon_{21}$ = $-1$. In general $u$ is complex valued, but a phase difference is trivially dealt with, so we make $u$ real.

	Taking a closer look at the eigenvector equation (1), $$ R v_{a} = [\sigma^{4} \cos (\theta / 2)+ i n^{k} \sigma^{k} \sin (\theta / 2)] v_{a} = [ \cos (\theta / 2) + i \sin (\theta / 2)] v_{a} = e^{i \theta/2} v_{a},$$ implies that $n^{k} \sigma^{k} v_{a}$ = $v_{a}.$ The point is that multiplication by either matrix $\sigma^{4}$ or $n^{k} \sigma^{k}$ does not change the eigenvectors $v_{1}$ and $v_{2}.$ Having two such matrices gives flexibility that is exploited to make the boosts $\Lambda^2$ and $\Lambda^{-2}$ of Fig.~1. 

	With $\sigma^{4}$ and $n^{k} \sigma^{k}$ we can generalize the numerical factor in (2) to a matrix expression,
\begin{equation} 
 v_{a} = (\sigma^{4} p_{ab4} - \epsilon_{ab} \sigma^{k} p_{ab} n^{k} ) v_{b},
\end{equation}
where 
\begin{equation} 
p_{ab4} = A + \epsilon_{ab} (B + \sinh u ) \hspace{0.5in} p_{ab} = B + \epsilon_{ab} ( A - \cosh u ).
\end{equation}
Here $A$ and $B$ are arbitrary. (2) is the special case $A$ = $\cosh u$ and $B$ = 0. Another special case gives the boosts $\Lambda^{\pm 2}$ of Fig.~1 which may be written as $\Lambda^{\epsilon_{ab} 2}$ = $\sigma^{4} \cosh u + \epsilon_{ab} \sigma^{k} n^{k} \sinh u.$ Thus we don't need to introduce the boosts $\Lambda$ of the Lorentz group and we can base the relationship between left and right spinors on the rotation group.

	Assumption I: $p_{ab4}$ and $p_{ab}$ are independent of $ab$. This makes the $\epsilon_{ab}$ coefficients in (4) vanish and we can write the coefficients as
\begin{equation} 
p_{4} \equiv p_{ab4} = A = \cosh u \hspace{1in} p \equiv  - p_{ab} =  - B =  \sinh u,
\end{equation}
so that $p >$ 0 when $u >$ 0. We set components $p^{k}$ = $-p_{k} $ = $n^{k} \sinh{u}.$ Thus the $p_{ab}$s take the $\cosh - \sinh$ form of a unit energy-momentum. If we call $v_{2}$ `LEFT' and $v_{1}$ `RIGHT', then the assumption says the `unit energy-momentum is the same going from right to left as going from left to right'.

	Equation (3) with Assumption I,
\begin{equation} 
 v_{a} = (\sigma^{4} \cosh u + \epsilon_{ab} \sigma^{k} \sinh u n^{k} ) v_{b},
\end{equation}
clones a relationship between the right and left spinors of a single particle state of a spin 1/2 particle. A standard derivation [1-3] uses boosts while we see the equation as an expression of the factor relating two simultaneous eigenvectors. 

	Calling the eigenvectors $v_{1}$ and $v_{2}$ `spin up', we can find two `spin down' eigenvectors with equivalent magnitudes and the same energy-momentum. Write $v_{1}$ and $v_{2}$ as $v_{1}^{+}$ = $ e^{u} \mid v_{2} \mid u^{+}$ and $v_{2}^{+}$ = $\mid v_{2} \mid u^{+}$. Then the spin down eigenvectors are $v_{1}^{-}$ = $ \mid v_{2} \mid u^{-}$ and $v_{2}^{-}$ = $e^{-u} \mid v_{2} \mid u^{-}.$ 

	To put (6) in a more commonly seen form, put $v_{1}$ together with $v_{2}$ in a four component column matrix $\psi \equiv$ col$\{v_{1}, v_{2} \}$. Then we find that (6) becomes
\begin{equation} 
\gamma^{\mu} p_{\mu} \psi= \psi , 
\end{equation}
where $\mu \in$ $\{1,2,3,4\}$ and the $\gamma$s are the $4 \times 4$ matrices
\begin{equation} 
 \gamma^{k} = \pmatrix{ 0 && - \sigma^{k} \cr \sigma^{k} && 0} \hspace{0.5in} \gamma^{4} = \pmatrix{ 0 && \sigma^{4} \cr \sigma^{4} && 0} . 
\end{equation}
These satisfy $\gamma^{\mu} \gamma^{\nu} + \gamma^{\nu} \gamma^{\mu}$ = 2 $I$ diag$\{-1,-1,-1,+1\}^{\mu \nu},$ where $I$ is the unit $4 \times 4$ matrix. We recognize the space-time metric and it follows that the $\gamma$s are Dirac matrices. With these $\gamma$s the Dirac equation is given in a `chiral' representation.

	A large void is the absence of space-time. Where are the coordinates that make up the variables conjugate to the energy-momentum? We turn now to that problem. 

\section{$\theta$ as Proper Time} 

	We envision $\theta$ as a possible proper time. Then we obtain quantities $dq^{\mu}$ where $d \theta \propto$ $p_{\mu} dq^{\mu}.$ The $dq^{\mu},$ where $\mu \in$ $\{1,2,3,4\},$ can then be the candidate coordinates for space-time. We also require that the gradient $\partial_{\mu} \equiv$ $\partial / \partial q^{\mu}$ be defined so that (6) can be rewritten as an operator equation. The gradient is chosen so that the operator equation is equivalent to the Dirac equation in a coordinate representation.
  
	By (1), applying the rotation $R$ with a variable rotation angle $\theta$ to the eigenvectors $v_{a}$ gives each the variable phase $\theta/2$,
\begin{equation} 
 v_{a}^{\prime} = e^{i(\theta/2)} v_{a},
\end{equation}
where $a \in$ $\{1,2\}.$ It is straightforward to verify the identity,
\begin{equation} 
 - 2 i \frac{d}{d \theta} v_{a}^{\prime} = v_{a}^{\prime}.
\end{equation}
Now we can rewrite (6) as 
\begin{equation} 
 v_{a}^{\prime} = -i[\sigma^{4}(2p_{4}\frac{d}{d \theta})  - \epsilon_{ab} \sigma^{k} (2p_{k}\frac{d}{d \theta})] v_{b}^{\prime},
\end{equation}
where $p_{k} \equiv$ $-p n^{k}$. We get an operator form of (6).

	We know what $\theta$ is; it is the rotation angle. The $q$s have no such foundation. So we are free to choose the $q$s any way we wish. Let us define the partials $\partial_{\mu}$ so that the operator version (11) takes a useful form. 

	In quantum mechanics the four-gradient $-i \partial_{\mu}$ occurs as the energy-momentum operator. Suppose, to start with, we define a gradient operation by
\begin{equation} 
 \frac{\partial}{\partial q^{\mu}} \equiv 2 p_{\mu} \frac{d}{d \theta}. \hspace{1in} {\mathrm{(First Try)}}
\end{equation}
Applying $\partial_{\mu}$ to $\theta$ we get,
\begin{equation} 
 \frac{\partial \theta}{\partial q^{\mu}} = 2 p_{\mu}.
\end{equation}
Since $\theta$ has constant partials, $\partial_{\mu} \theta $ = $2 p_{\mu}$, it follows that $d \theta$ = $2 p_{\mu} dq^{\mu}$ and $\theta_{2} - \theta_{1}$ = $2 p_{\mu} (q_{2}^{\mu} - q_{1}^{\mu}).$ Thus assuming (12) implies $\Delta \theta$ is the phase of a plane wave; in particular $\Delta \theta$ is a single-valued function of $\Delta q^{\mu}.$ These are possible coordinates, but there are other choices. See Fig.~3.
\begin{figure}[h] 	
\vspace{0in}
\hspace{0.75in}\includegraphics[0,0][360,180]{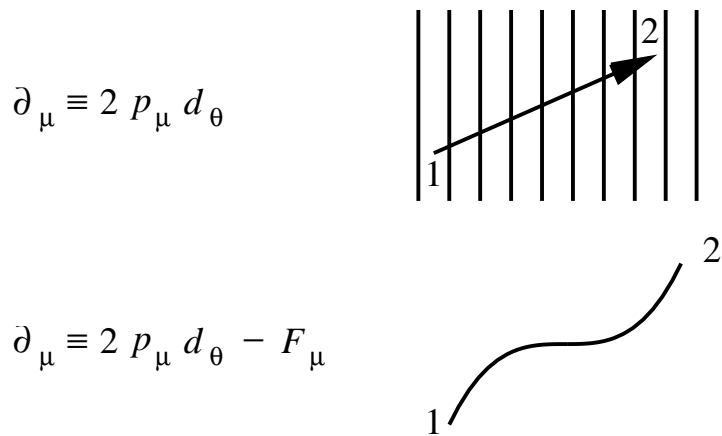}
\caption{The definition of the space-time gradient determines how the rotation angle $\theta$ depends on position in space-time. For the upper definition, the map of $\theta$ makes a plane wave in which the change of $\theta$ from space-time event 1 to event 2 is the twice the inner product of the interval 12 with $2p_{\mu},$ $\Delta \theta$ = $2 p_{\mu} \Delta q^{\mu}.$ More generally, for the gradient defined with four functions of position $F_{\mu}$, the rotation angle depends on the path from 1 to 2. }
\end{figure}

	Requiring $\theta$ to be a single-valued function of the $q$s is not the only possibility. When we assume (12) we get a plane wave so we say we are describing a `free' particle. When the difference between $\partial_{\mu} $ and $2 p_{\mu} (d / d \theta)$ is a single-valued function of the $q$s, then we describe a `particle interacting with a field described by a potential' that is a single-valued function of the $q$s.

	Assumption II. Let the four functions $F_{\mu}(q)$ be the differences between $2 p_{\mu} (d / d \theta)$ and $\partial_{\mu}.$ Then we define the partials
\begin{equation} 
 \frac{\partial}{\partial q^{\mu}} \equiv 2 p_{\mu} \frac{d}{d \theta} - F_{\mu}. 
\end{equation}
We must be much more careful with (14) than we were with (12). Consider applying (14) to a function $f(\theta),$ assuming that $\theta$ is a function of the $q$s, $\theta (q^{\mu}).$ Then one finds that, for $d_{\theta} f$ = $df / d \theta \neq$ 0,
$$ \frac{\partial \theta}{\partial q^{\mu}} = 2 p_{\mu} - F_{\mu} \frac{f}{(df / d \theta)} \hspace{0.5in} ( = \frac{\partial_{\mu} f}{d_{\theta} f} ).$$
The partial derivatives of $\theta$ depend on the function $f$, except for the free particle case $F_{\mu}$ = 0. Thus we cannot apply the operator (14) to all functions $f$ without contradictions arising. However, we only need to apply (14) to one function since the identity (10) is true for the function $e^{(i \theta /2)}$ in (9) and not true for other functions of $\theta.$

	By (9) the function $f$ is restricted to $f$ = $e^{(i \theta /2)}$ and it follows that 
\begin{equation} 
 \frac{\partial \theta}{\partial q^{k}} = 2 p_{\mu} + 2 i F_{\mu} \hspace{0.5in} {\mathrm{and}} \hspace{0.5in}  \theta = 2 \int ( p_{\mu} + i F_{\mu}) dq^{\mu} .
\end{equation}
Thus, in general, the phase is a path-dependent function of position $q^{\mu}$ with the gradient defined by (14). [If $F_{\mu}$ = $\partial_{\mu} g$ throughout some region, then the phase is path independent within that region.] Path-dependent phase explains many experimental results, e.g. the interference pattern made by electrons flowing past a solenoid. In such cases the functions $F_{\mu}$ are related to the electromagnetic vector potential.

	Let the functions $ F_{\mu}$ be proportional to the `electromagnetic vector potential' $ A_{\mu},$ $F_{\mu}$ = $ - i e A_{\mu},$ where $e$ is the `charge'. The operator equation (11) becomes
\begin{equation} 
 v_{a}^{\prime} = -i[\sigma^{4}(\frac{\partial}{\partial q^{4}} - i e A_{4})  - \epsilon_{ab} \sigma^{k} (\frac{\partial}{\partial q^{k}} - i e A_{k})] v_{b}^{\prime}.
\end{equation}
Just as we did with (6), we rewrite (16) with $\psi$ now made with  $v_{1}^{\prime}$ and $v_{2}^{\prime},$  
\begin{equation} 
 -i\gamma^{\mu}(\frac{\partial}{\partial q^{\mu}} - i e A_{\mu})\psi = \psi.
\end{equation}
Other rotations and eigenvectors satisfy this same equation since the gradient produces the appropriate value of $p_{\mu}$ for each. If we assume all these occur with the same $A_{\mu}$s then linear combinations of $\psi$s satisfy (17) and we have the superposition principle. In particular the `spin up' eigenvectors of Fig.~1 have a matching pair of spin down eigenvectors whose eigenvalues have negative the phase in (9). Linear combinations of spin up and spin down $\psi$s can make any spin direction. It follows that (17) is indeed the Dirac equation and can be interpreted to describe a charged spin 1/2 particle with unit mass. 


\appendix
\section{Problems}

1. Use the spin matrices displayed after equation (1) to calculate $n^{k} \sigma^{k},$ where summation over $k$, $n^{k} \sigma^{k}$ = $n^{1} \sigma^{1}$ + $n^{2} \sigma^{2}$ + $n^{3} \sigma^{3},$ is understood. Show that $(n^{j} \sigma^{j}).(n^{k} \sigma^{k})$ = $\sigma^{4}$ when ${(n^{1})}^2$ + ${(n^{2})}^2$ + ${(n^{3})}^2$ = 1.

\noindent 2. The notation $R$ = $e^{(i n^{k} \sigma^{k} \theta/2)}$ means $$R = \sigma^{4} + i n^{k} \sigma^{k} \theta/2 + \frac{(i n^{k} \sigma^{k} \theta/2)^2}{2} + \ldots ,$$ where the series is the expansion of the exponential function. Use the result in Problem 1 to show that $e^{(i n^{k} \sigma^{k} \theta/2)} = \sigma^{4} \cos (\theta / 2)+ i n^{k} \sigma^{k} \sin (\theta / 2).$ 

\noindent3. Let $\rho$ and $\phi$ be the polar and azimuthal angles for $n^{k}$, $n^{k}$ = $( \sin \rho \cos \phi, \sin \rho \sin \phi, \cos \rho).$ Show that $$u^{+} = \pmatrix{ \cos{(\rho/2)} e^{(-i\phi/2)} \cr \sin{(\rho/2)} e^{(+i\phi/2)}}$$ is an eigenvector of the rotation matrix $R$ = $e^{(i n^{k} \sigma^{k} \theta/2)}.$ Also find a $u^{-}.$

\noindent4. Suppose we introduce four parameters in two four-vectors $j^{\mu}$ = $\{j n^{k}, j^{4}\}$ and $a^{\mu}$ = $\{a n^{k}, a^{4}\}$ and rewrite (4) in the form
$$ p_{ab4} = \cosh{u} + K j^{4} + \epsilon_{ab} K a^{4} \hspace{0.5in} p_{ab} = - \sinh{u} + K j - \epsilon_{ab} K a. $$
Show that (2) and (3) imply $j$ = $a^{4},$ $j^{4}$ = $a,$ and $j^{4} a^{4} - j^{k} a^{k}$ = 0. The scale factor $K$ is arbitrary. What kind of interaction might this describe if $j^{\mu}$ and $a^{\mu}$ are the current and spin vectors of a source spin 1/2 particle? What value should $K$ have for electroweak interactions? What effects would the $a^{\mu}$ term have?

\noindent5. Let $F_{\mu}$ in (14) be linear in the gradients $\partial_{\mu}$: i.e. $$F_{\mu} \rightarrow F_{\mu} + M_{\mu \nu} \partial_{\nu},$$
where M is independent of $q_{\mu}.$ Find the new dependence of the rotation angle $\theta$ on position $q_{\mu}.$

\end{document}